\documentclass[preprint,superscriptaddress]{revtex4-1}

\usepackage{graphicx}
\usepackage{dcolumn}
\usepackage{bm}
\usepackage{amsmath, amsthm, amssymb}
\usepackage{xcolor}

\newcommand{\be}{\begin{equation}}
\newcommand{\ee}{\end{equation}}
\newcommand{\bea}{\begin{eqnarray}}
\newcommand{\eea}{\end{eqnarray}}

\newcommand{\bef}{\begin{figure}}
\newcommand{\enf}{\end{figure}}

\begin{document}
\draft

\title{Robust transverse structures in rescattered photoelectron wavepackets and their observable consequences}


\author{Timm Bredtmann}
\email{Timm.Bredtmann@mbi-berlin.de}

\author{Serguei Patchkovskii}
\affiliation{Max-Born-Institute, Max-Born-Stra{\ss}e 2A, D-12489 Berlin, Germany}


\begin{abstract}
Initial-state symmetry has been underappreciated in strong-field
spectroscopies, where laser fields dominate the dynamics. We demonstrate
numerically that the transverse photoelectron phase structure, arising from
this symmetry, is robust in strong-field rescattering, and manifests in
strong-field photoelectron spectra. Interpretation of rescattering experiments
need to take these symmetry effects into account. In turn, transverse
photoelectron phase structures may enable attosecond super-resolution imaging
with structured electron beams.
\end{abstract}
\maketitle

Symmetries, exact and approximate, and symmetry breaking underpin our
understanding of perturbative
spectroscopies~\cite{bunker_fundamentals,bunker_symmetry}, where field-free
symmetries of initial  and final states constrain the range of outcomes.  In
contrast, in intense infrared laser fields, the dynamics are dominated by the
field itself.  For example, strong infrared pulses can remove an electron from an
atom, a molecule or in a solid (ionization), accelerate it (propagation), and
finally drive it into the parent ion (recollision).  This is the three-step
model~\cite{kulander,corkum}, capturing the essence of the emerging field of
attosecond (asec) strong-field
spectroscopies~\cite{krausz_ivanov,lin_rescattering_review,vrakking_attosecond}.
In the recollision step the returning electronic wavepacket (REWP) may
recombine to the parent-ion, giving rise to high-harmonics
generation~\cite{krausz_ivanov,lin_rescattering_review,vrakking_attosecond,lewenstein_sfa,olga_sfa,qrs,lin_rescattering_review},
or scatter elastically, yielding strong-field photoelectron
holography~\cite{huisman,spanner_nat_physics_2014,woerner_nat_com_2017} and
laser-induced electron diffraction
(LIED)~\cite{krausz_ivanov,lin_rescattering_review,qrs,lied1,LEIN_lied,lied2,lied_meckel,lied_blaga,fables_xu,lied_biegert,fables_biegert,Ueda_lied,lied_wolter_biegert,woerner_nat_com_2017}.

The symmetry of the laser field, and the dynamical symmetry it imposes on the
continuum wavepacket, determine many qualitative features of these
processes by means of selection rules \cite{moiseyev_selection_rules}. It is widely
recognized that tunnel-ionized electrons that do not recollide give information
about the initial-state symmetry \cite{lied_meckel,woerner_nat_com_2017}.  At
higher energies, holographic patterns are sensitive to the phase-structure
of the REWP~\cite{spanner_nat_physics_2014}, which may arise from the asymmetry
in the binding-potential~\cite{spanner_nat_physics_2014} or the initial-state
symmetry~\cite{woerner_nat_com_2017}. However, in other strong-field
spectroscopies such as LIED it is commonly assumed that the laser field
entirely dominates subsequent dynamics, and the transverse phase structure due
to the initial-state symmetry is ``washed'' out in the propagation
step~\cite{qrs}, yielding an asymptotically-flat wavefront. This
assumption is implicitly enforced by the stationary-phase treatment of the
strong-field approximation (SFA)~\cite{lewenstein_sfa,olga_sfa}. 


Recently, an experimental/theoretical study in our institute~\cite{schell_sci_adv_2018} has shown
that the rescattering probability in trans-butadiene is specific to the
ionization channel and the molecular orientation, rather than a property of the
driving field alone.  These results imply that the REWP may retain, despite the
strong driving field, the transverse phase structure imprinted by the initial
state.  Such structure will significantly influence the recollision process and
alter the shape and the interpretation of strong-field photoelectron spectra.
Ultimately it may enable atomic-scale engineering of structured electron beams
-- an electron-beam analogue to structuring the illumination in
super-resolution light microscopy~\cite{Hell_Science_review}.

Unfortunately, limited statistics~\cite{schell_sci_adv_2018} precluded
measuring the angle- and ionization-channel-resolved photoelectron spectra,
permitting alternative interpretations. The goal of this Letter is to numerically explore
the consequences of the initial-state symmetry for the LIED region of
strong-field photoelectron spectra.  We are particularly interested in
determining whether these effects are robust to the misalignment between the
field-free symmetry elements and the laser-field polarization.


In molecular systems, the initial-state symmetry effects are intertwined with
contributions due to the potential asymmetry \cite{spanner_nat_physics_2014},
orientation, and nuclear motion. Combined with the cost of solving the
time-dependent Schr\"odinger equation (TDSE), this complexity makes a
conclusive analysis challenging. Instead, we consider the simplest-possible
example of the initial-state symmetry: a one-electron atom (He$^+$ or
``Argon''~\cite{tsurff_spherical}), initially in an antisymmetric $p$ state.
We solve the three-dimensional TDSE in the velocity gauge and dipole
approximation~\cite{tsurff_spherical,tdse_spherical}. 
The simulation grid contains 2200 radial points with
spacings of $\Delta{}r=0.2$ $a_0$, and angular momentum channels up to
$L\leq{}40$, $|M|\leq{}40$. The time-step is $\Delta{}t=0.03\,\rm{asec}$ and a
transmission-free absorbing potential~\cite{cap} is applied at $r=407.6$ $a_0$.
The photoelectron spectra are calculated using surface-flux
integration~\cite{PhysRevA.60.4831,tsurff_scrinzi} continued to infinite
time~\cite{tsurff_spherical}.
Sine-squared envelopes are used for the vector
potential $A(t)$, with carrier-envelope phase of $\pi/2$ (corresponding to the
time-odd electric field $E(t)=-\partial A(t)/\partial t$).

\begin{figure}[h]
  \centering
     \includegraphics[width=0.5\columnwidth]{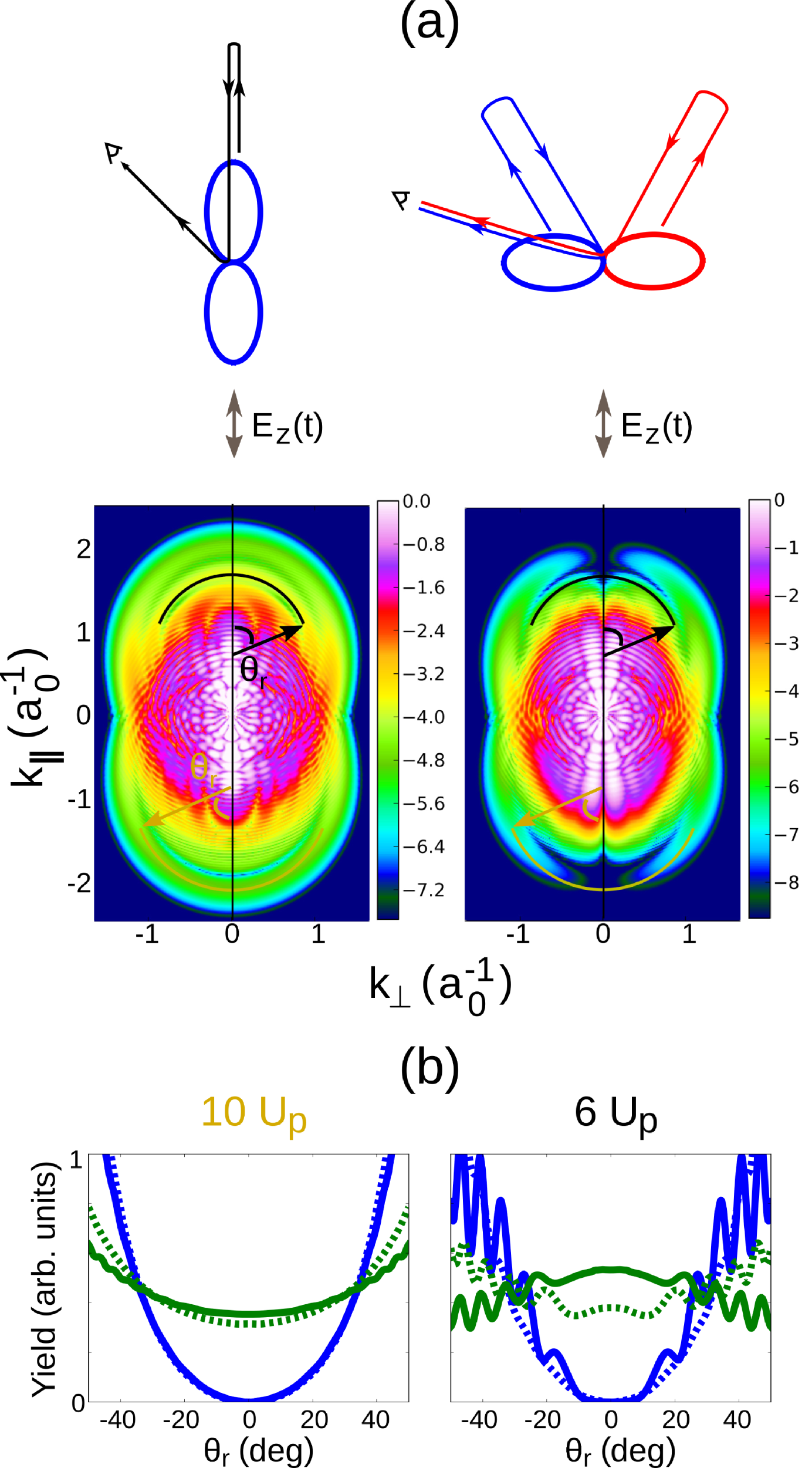}
\caption{
(a) Angle-resolved strong-field photoelectron spectra for the He$^+$ ion for
initial states symmetric ($|2p_z\rangle$, left) and anti-symmetric
($|2p_x\rangle$, right) with respect to the laser polarization direction.  The
strong ($10^{14}$ W/cm$^2$) 5-cycle, 800 nm driving field is linearly polarized
along the $z$ axis.  The 10U$_{\rm{p}}$ and the 6U$_{\rm{p}}$ recollision circles are
indicated by the yellow (upper) and the black (lower) lines, respectively, with the definition of the deflection
angle $\theta_r$. The ionization probabilities are 0.16
($|2p_z\rangle$) and 0.12 ($|2p_x\rangle$). The spectra are shown on a
logarithmic scale.  (b) Angle-resolved photoelectron yields along the two
recollision circles for $|2p_z\rangle$ (green, upper lines) and $|2p_x\rangle$ (blue, lower lines).  Left
(right) side: yields along the 10U$_{\rm{p}}$ (6U$_{\rm{p}}$) circle.  All
curves are normalized to constant area. Dashed lines:
Focal-averaged yields at peak intensity of $10^{14}$ W/cm$^2$; see text and
appendix for details.
}
\label{Fig1}
\end{figure}

Figure~\ref{Fig1}(a) shows strong-field photoelectron spectra for the
He$^+$ ion.  The 5 cycle, 800 nm electric field with peak intensity
$I_{\rm{max}}=10^{14}$ W/cm$^2$ is linearly-polarized along the $z$ direction.
At first, we explore two initial states
$|\Psi_0\rangle$: symmetric ($|\Psi_0\rangle=|2p_z\rangle$, left column) and
anti-symmetric ($|\Psi_0\rangle=|2p_x\rangle$, right column) with respect to
the $\sigma_{yz}$-reflection. The vertical axis ($k_{\parallel}$) coincides
with the laser polarization direction and the horizontal axis ($k_{\perp}$)
with the ``averaged'' perpendicular direction.
The averaging of the strong-field photoelectron spectra is performed in spherical coordinates over the
azimuthal angle $\phi$, separately in the ``left'' and ``right'' hemispheres.
Hence, the ``left'' side of the spectra ($k_{\perp}<0$) average over
$90^{\circ}<\phi<270^{\circ}$; the ``right'' side ($k_{\perp}>0$) over $\phi<90^{\circ}$ and $\phi>270^{\circ}$.
Finally, the polar angle $\theta$ is defined with respect to the laser polarization
direction ($k_{\parallel}$-axis).  


For the symmetric $2p_z$ case (left panel), we observe a typical strong-field
photoelectron spectrum symmetric with respect to perpendicular momenta
$k_\perp$, including holographic ``fingers'' in the lower-energy region below
2U$_{\rm{p}}$ (U$_{\rm{p}}$ is the ponderomotive energy)
\cite{huisman,spanner_nat_physics_2014,woerner_nat_com_2017} and recollision
circles for back-scattered electrons \cite{qrs} (small deflection angles
$\theta_r$ in our notation). The 10U$_{\rm{p}}$ (6U$_{\rm{p}}$) recollision
circles and the definition of the deflection angle $\theta_r$ are indicated by
the yellow (black) lines.  The 10U$_{\rm{p}}$ circles correspond to electrons
that return with maximal kinetic energy to the ion core, elastically
backscatter ($|\theta_r|<90^{\circ}$), and gain an additional drift momentum
equal to the vector potential at the moment of recollision $A(t_r)$~\cite{qrs}.
The maximal energy gain corresponds to $\theta_r=0^{\circ}$.  This definition
of the deflection angle is chosen to simplify the discussion of orientational
averaging below. It differs from the conventional electron-diffraction
definition~\cite{qrs} by $180^\circ$.  For the anti-symmeric $2p_x$ case (right
panel), i.e. polarization of the laser field within the symmetry plane, the
photoelectron spectra change dramatically, now vanishing for $k_{\perp}=0$
($\theta_r=0^{\circ}$).


The corresponding angle-resolved spectrum along the 10U$_{\rm{p}}$
(6U$_{\rm{p}}$) circle is shown in the left (right) panel of
Fig.~\ref{Fig1}(b) (solid lines). Dashed lines give the
focal-spot average (see appendix for details, including the full angle-resolved 2D maps in Fig. S5). For the symmetric
case (($|2p_z\rangle$, green line), the yield along the 10U$_{\rm{p}}$ circle
largely follows the Rutherford scattering cross-section, allowing retrieval of
structural information in LIED~\cite{qrs}. For the anti-symmetric case
($|2p_x\rangle$, blue line), the yield for the perfect back-scattering
($\theta_r=0^{\circ}$) vanishes and its angular dependence is changed
qualitatively. Focal averaging does not significantly affect
these results. The 6U$_{\rm{p}}$ circle (right) shows the same characteristic
difference between the symmetric (green) and anti-symmetric (blue) case,
although the detailed structure is complicated due to intensity-dependent
interferences with late electron
returns \cite{krausz_ivanov,off_axis_milosevic}. These interferences lead e.g. to a maximum at $\theta_r=0^{\circ}$
for the anti-symmeric case (blue solid line) and 
are largely suppressed by focal averaging (dashed lines).

These results agree with the SFA rescattered-photoelectron
amplitude~\cite{abecker_sfa}:
\begin{equation}
a({\bf k}_{f})=-\int{\rm d}t_{0}{\rm d}t_{c}{\rm d}{\bf p}e^{-iS_{V}(t_{0},t_{c},{\bf p})}\times{}R\times{}I,
\label{eq_sfa1}
\end{equation}
where the elastic scattering and photoionization matrix elements are:
\begin{align}
 R&=\langle{\bf k}_{f}+A_z(t_{c})|V_{c}|{\bf p}+\hat{\bf z}A_z(t_{c})\rangle, \nonumber \\
 I&=\langle{\bf p}+A_z(t_{0})|V_{L}(t_{0})|\Psi_{0}\rangle. \nonumber
\end{align}
Integration in Eq.~\eqref{eq_sfa1} is over the ionization and recollision-times
$t_{0}$ and $t_{c}$, respectively, and the canonical momentum ${\bf p}$. The
quantity $S_{V}(t_{0},t_{c},{\bf p})$ is the length-gauge Volkov
phase~\cite{olga_sfa} and $\hat{\bf z}A_z(\tau)$ the vector potential
($\hat{\bf z}$ is the unit vector in $z$ direction).  Finally, $V_{c}$ is the
Coulomb potential, $V_{L}(t_{0})=zE_z(t_{0})$ describes the interaction of the
$z$-polarized laser field $E_z(t_0)$ and the atom, and $|\Psi_0\rangle$ is the
initial state. 

In the lowest-order stationary-phase approximation (SPA) Eq.~\eqref{eq_sfa1}
becomes~\cite{abecker_sfa}:
\begin{equation}
a({\bf k}_{f})\propto{}e^{-iS_{V}(t_{0s},t_{cs},{\bf p}_{s})}\times{}R_s\times{}I_s,
\label{eq_sfa2}
\end{equation}
with the stationary matrix elements
\begin{align}
 R_s &=\langle{\bf k}_{f}+A_z(t_{cs})|V_{c}|\hat{\bf{z}}(p_{sz}+A_z(t_{cs}))\rangle \nonumber \\
 I_s &=\langle \hat{\bf{z}}(p_{sz}+A_z(t_{0s}))|V_{L}(t_{0s})|\Psi_0\rangle \nonumber,
\end{align}
where $t_{0s}, t_{cs}$ and ${\bf p}_{s}$ are the stationary points of the
Volkov phase. The perpendicular component of
the stationary canonical momentum $p_{s\perp}$ vanishes~\cite{abecker_sfa,olga_sfa}.

For the symmetric $2p_z$ case, $R_s$ is the Rutherford scattering
cross-section~\cite{qrs} and the recollision is described by a single
trajectory with zero transverse momentum (Fig.~\ref{Fig1}(a), upper-left
panel). For the anti-symmetric $|2p_x\rangle$ initial state the stationary
hydrogenic photoionization matrix element:
\begin{equation}
 I_s \propto\frac{(p_{sz}+A_z(t_{0s}))p_{sx}}{(4(p_{sz}+A(t_{0s}))^2+4)^4)} \nonumber
\end{equation}
vanishes by symmetry ($p_{sx}=0$). The lowest-order SPA rescattered amplitude
then vanishes \cite{wilhelm_sfa}, and higher orders must be considered in
evaluating Eq.~\eqref{eq_sfa1}.  Expanding the matrix elements $I$ and $R$ in
Eq.~\eqref{eq_sfa1} in the transverse component of the canonical momentum
$\bf{p}$ around the stationary point ${\bf p}_s$, we obtain, in the lowest
surviving order~\cite{schell_sci_adv_2018}:
\begin{align}
a({\bf k}_{f}) & \propto e^{-iS_{V}(t_{0s},t_{cs},{\bf p}_{s})}\left(\frac{1}{i(t_{cs}-t_{0s})}\right)\times{}\tilde{R}_s\times\tilde{I}_s
\label{eq_sfa3} \\
\tilde{I}_s & = \frac{\partial}{\partial p_{x}}\langle{\bf p}+A_{z}(t_{0})|V_{L}(t_{0})|2p_{x}\rangle\Bigg\vert_{{\bf p}=p_{sz}} \nonumber \\
\tilde{R}_s & = \frac{\partial}{\partial p_{x}}\langle{\bf k}_{f}+A_{z}(t_{c})|V_{c}|{\bf p}+A_{z}(t_{c})\rangle\Bigg\vert_{{\bf p}=p_{sz}} \nonumber.
\end{align}
For a hydrogenic state:
\begin{align}
\tilde{I}_s & \propto\frac{(p_{sz}+A_z(t_{0s}))}{(4(p_{sz}+A(t_{0s}))^2+4)^4)} \nonumber \\
\tilde{R}_s & \propto \frac{k_{fx}}{(|{\bf k}_f - p_{sz}|^2)^2}. \nonumber
\end{align}

Thus, for laser polarization along a symmetry plane, the REWP can no longer be
described by an asymptotically-flat wavefront with a well-defined return
direction along the laser polarization. Instead, we can introduce an
(arbitrary) small transverse momentum $\Delta p_{sx}$.  For an atomic $2p_x$
initial state, the finite-difference expression for $\tilde{I}_s$ and
$\tilde{R}_s$ then factorizes into a sum of two contributions (or trajectories)
with $\pi$ phase difference, added coherently (Fig.~\ref{Fig1}(a), top right).  
Their interference alters the high-energy (LIED) region of the 
photoelectron spectrum, in particular suppressing the signal for $k_{fx}=0$.


\begin{figure}[h]
  \centering
     \includegraphics[width=0.7\columnwidth]{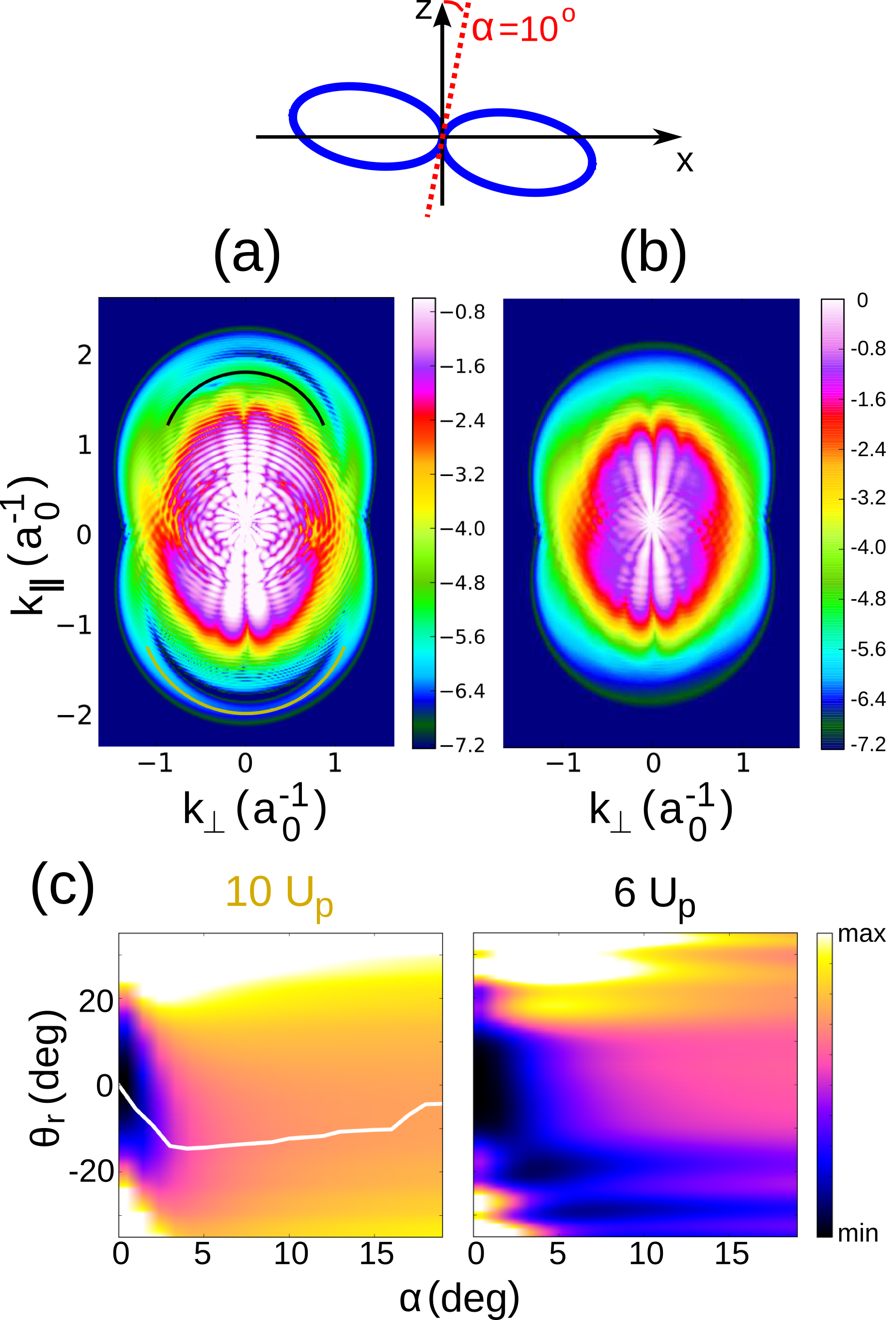} \\

\caption{(a) Strong-field photoelectron spectrum for a ``misaligned'' He$^+$
ion.  The initial state $|\Psi_0\rangle$ is a 2p$_x$ orbital rotated by
$\alpha=10^{\circ}$ around the $y$ axis. The spectrum is shown on a logarithmic scale. (b) Same as (a),
including focal-spot averaging. (c) Angle-resolved photoelectron yield along
the 10U$_{\rm p}$ and 6U$_{\rm p}$ recollision circles, as function
of the misalignment angle $\alpha$ for the single-intensity
case (panel a). For the 10U$_{\rm{p}}$ circle (left), the solid line traces the minimum
for each $\alpha$.  The yields are normalized to the same area for each
$\alpha$.  See Fig.~\ref{Fig1} caption for the field parameters and other
definitions.  }
\label{Fig2}
\end{figure}

In order to explore the robustness of our results with respect to misalignment of the laser-polarization
direction and the symmetry plane of the initial state, Figure~\ref{Fig2}(a) shows
the photoelectron spectrum for a misalignment angle of $\alpha=10^{\circ}$.
The same $z$ polarized
pulse as in Fig.~\ref{Fig1}(a) with single peak intensity $I_{\rm{max}}=10^{14}$ W/cm$^2$ is used.
Moreover, Fig.~\ref{Fig2}(b) shows the corresponding focal-averaged spectrum with the same peak intensity.
Both in the single-intensity and in the focal-averaged photoelectron spectrum
 the signatures of the phase structure of
the REWP are quite
different in the lower-energy (holographic) and the higher-energy (LIED)
region:
In the holographic region, the photoelectron signal is suppressed close to $k_{\perp}=0$, clearly reflecting the symmetry plane of the
initial state~\cite{woerner_nat_com_2017}.  This behaviour is qualitatively
similar to exact laser polarization along a symmetry plane ($\alpha=0^{\circ}$) (Fig.~\ref{Fig1}(a), right panel).
As expected, a ``left''--``right''
asymmetry of the photoelectron signal is present for $\alpha=10^{\circ}$,
which breaks the overall reflection symmetry.

In the LIED region effects of the initial-state symmetry
are more subtle for $\alpha=10^{\circ}$.
Here, the non-symmetric initial state gives rise to a transverse phase gradient of the REWP.
This phase gradient results in a left--right
asymmetry, and the corresponding displacement of the minimum along the
recollision circles.
This behaviour is illustrated further in Fig.~\ref{Fig2}(c) which shows the angle-resolved photoelectron yield for the single-intensity case (Fig.~\ref{Fig2}(a))
along the same recollision-circles
used in Fig. \ref{Fig1} as function of the misalignment angle $\alpha$.
For the 10U$_{\rm{p}}$ circle (left panel), the white line traces
the photoelectron minimum for each $\alpha$. For $\alpha=10^{\circ}$ the
minimum lies around $\theta_r=-13^{\circ}$. This displacement deviates
considerably from the Rutherford scattering cross section with minimum at
$\theta_r=0^{^\circ}$ (Fig.~\ref{Fig1}(b), left panel).
This behaviour corresponds to counter-rotation of the photoelectron signal
with the initial state which prevails well beyond
$\alpha=20^{\circ}$ (results are shown up to $\alpha=20^{\circ}$). The
counter-rotation is even more pronounced along the 6U$_{\rm{p}}$ recollision circle
(Fig.~\ref{Fig2}(c), right panel).
\begin{figure}[t]
  \centering
     \includegraphics[width=0.77\columnwidth]{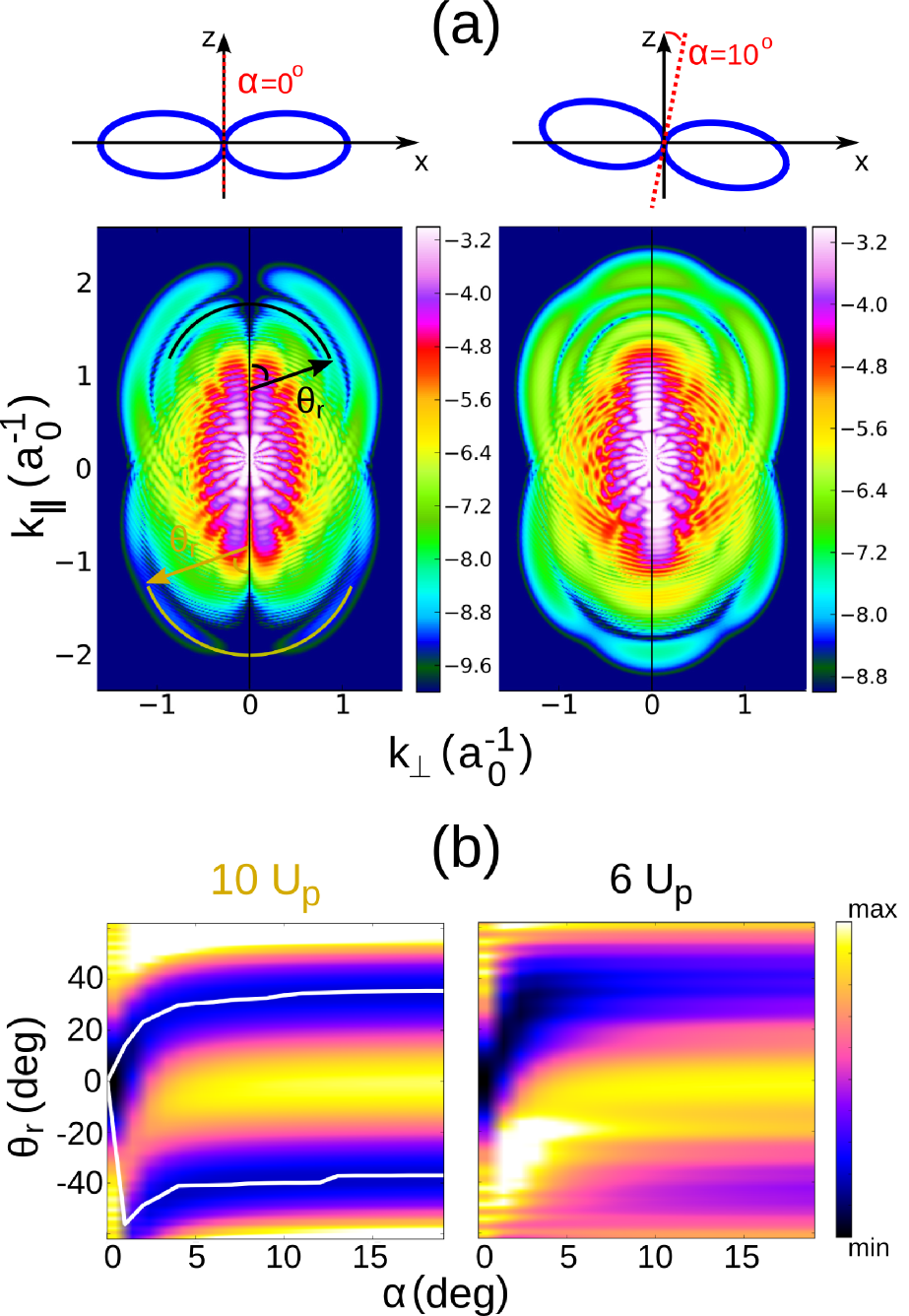}
\caption{(a) Strong-field photoelectron spectra for the model argon atom (see
text) for two different initial states: $|\Psi_0\rangle=|3p_x\rangle$ (left)
and a $3p_x$ orbital rotated by $\alpha=10^{\circ}$ (right).  (b)
Angle-resolved photoelectron yield along the 10U$_{\rm p}$ and 6U$_{\rm p}$ recollision
circles as a function of the misalignment angle $\alpha$. See Fig.~\ref{Fig1}
caption for field parameters and further details.}
\label{Fig3}
\end{figure}
This left--right asymmetry can be understood as the interference of the signals
from equations~\eqref{eq_sfa2} and \eqref{eq_sfa3}, sketched
in Fig.~\ref{Fig1}(a).  With
$|\Psi_0\rangle=\rm{sin}(\alpha)\,|2p_z\rangle+\rm{cos}(\alpha)\,|2p_x\rangle$,
we get:
\begin{equation}
a({\bf k}_{f}) \propto e^{-iS_{V}(t_{0s},t_{cs},{\bf p}_{s})} \times 
 \left({\rm sin}(\alpha)\,R_{s}\times I_{s}+{\rm cos}(\alpha)\left(\frac{1}{i(t_{cs}-t_{0s})}\right)\tilde{R}_{s}\times\tilde{I}_{s}\right).
\label{eq_sfa4} 
\end{equation}
(In the holographic region, an additional interference with the direct-electron
``reference'' wave will also be present.)
Hence, also for an imperfect alignment the photoelectron phase structure
clearly influences the recollision process both in the
holographic and LIED region.
These results are robust with respect to focal averaging (see Figure S5).

%
%

Finally, Figure~\ref{Fig3}(a) shows the photoelectron spectrum for an ``Argon''
atom interacting with the same single-intensity laser field used in Figs.~\ref{Fig1} and
\ref{Fig2}.  For laser polarization exactly along a symmetry plane
($|3p_x\rangle$; left panel), results are similar to the He$^+$ ion (cf.
Fig.~\ref{Fig1}(a), right).  In particular, the photoelectron signal vanishes for $k_{\perp}=0$
due to the symmetry plane of the initial state, both in the holographic and the
LIED regions.  For $\alpha=10^{\circ}$ (right panel), significant differences
between the argon atom and the He$^+$ ion occur: In the holographic region,
the photoelectron spectrum co-rotates with the initial state for the argon
atom, whereas it is strongly suppressed
around $k_{\perp}=0$ for the He$^+$ ion (Fig.~\ref{Fig2}(a)).

In the LIED region the symmetry-induced phase structure of the REWP also
results in co-rotation of the signal for $\alpha\neq{}0^{\circ}$, and the
corresponding shift of the minima along the recollision circles.  For the
10U$_{\rm{p}}$ circle (Fig.~\ref{Fig3}(b), left), white lines trace the
photoelectron minima for each $\alpha$. For $\alpha=10^{\circ}$, the minima
along $\theta_r$ are shifted by $\approx4^{\circ}$ relative to the symmetric
case ($\alpha=90^{\circ}$).  As for the He$^+$ ions, the left--right asymmetry is more pronounced along
the 6U$_{\rm{p}}$ recollision circle (Fig.~\ref{Fig3}(b), right). 
However, we observe a co-rotation of the
photoelectron signal with the initial state, whereas the signal
counter-rotates with the initial state for the He$^+$-ion (Fig.~\ref{Fig2}(c)), illustrating the sensitivity of the phase structure of
the REWP on the specific quantum system.

%
%


In summary, we demonstrate that the initial-state symmetry imposes a phase
structure on the recolliding electron wavepacket, which modifies strong-field
rescattering.  
Thus, the complete characterization of the
rescattered photoelectron may be possible not only in the holographic
\cite{spanner_nat_physics_2014}, but also in the background-free
high-energy rescattering region of the strong-field photoelectron spectrum.
The signatures of the photoelectron`s phase structure,
especially of its transverse phase gradient,
are robust in the strong-field spectra to field misalignment with respect to the symmetry plane and
depend sensitively on the quantum system (Figs.~\ref{Fig2} and \ref{Fig3}).
Moreover, these signatures are robust with respect to focal averaging (Figs. \ref{Fig1}(b) and \ref{Fig2}(b))
 and with respect to the peak laser intensity (appendix).
Hence, initial-state symmetry needs to be accounted for in the
interpretation of strong-field recollision experiments both in the lower-energy
(holographic) and the higher-energy (LIED) region.  The numerical results are
supported by an analytical model based on an extension of the standard
stationary-phase approximation of the rescattered-electron
SFA~\cite{lewenstein_sfa,olga_sfa,schell_sci_adv_2018}. While
we concentrate on the holographic and the LIED regions, we expect that other
features of strong-field photoelectron spectra such as interference
carpets \cite{carpets} may be affected as well. 

The symmetry-induced continuum phase structures are expected to be important
for complex molecular systems as well, by interchanging the
structural minima and maxima in laser-induced diffraction (see \cite{lein_hhg_minima} for a similar effect in high harmonics generation 
and the appendix).  While strong-field ionization along a nodal
plane may be suppressed in small molecules~\cite{muth_sfi_suppression}, this
constraint is relaxed for a broad range of typical organic
molecules~\cite{schell_sci_adv_2018,spanner_bhardwaj_hhg_ellipticity},
where coincidence measurements may be used to disentangle
contributions from different ionization channels \cite{schell_sci_adv_2018}.

In turn, robust transverse phase structure of the returning wavepacket is the
prerequisite for attosecond super-resolution imaging with structured electron
beams.  For example, initial states carrying
ring-currents~\cite{barth_ring_current_jacs,doerner_ring_current} would lead to
atomic-scale electronic vortex (doughnut) beams, analogous to light vortices
used in super-resolution light microscopy~\cite{Hell_Science_review}.  Such
structured attosecond electron beams may specifically probe certain regions
within a molecule and thereby enhance the spatial resolution. While the
detailed investigation of nano-structured electron beams will be presented
elsewhere \cite{in_preparation}, the appendix gives illustrative
examples.  We show how the phase structure of such beams, coming from the
initial $p_{\pm}$, $d_{\pm}$, and real-valued $d$ states, are mapped,
background-free, onto the angle-resolved rescattering spectra (Figs. S1, S2).
In molecules, the transverse phase gradient of the nano-structured returning
electron beam is expected to enhance interferences between individual scattering centers
(atoms), resolving structural features well-below the usual de~Broglie limit
(Figs. S3, S4).  Conversely, a known molecular structure can also be used to
characterize the phase structure of the returning electronic wavepacket.


\begin{acknowledgments}
We thank Jens Biegert, Jochen Mikosch, and Michael Spanner for inspiring discussions and helpful advice.
\end{acknowledgments}

\renewcommand{\thefigure}{S\arabic{figure}}
\setcounter{figure}{0} 

\appendix
\section{Initial-state symmetries}
\begin{figure}[h]
\centering
\includegraphics[width=0.9\textwidth]{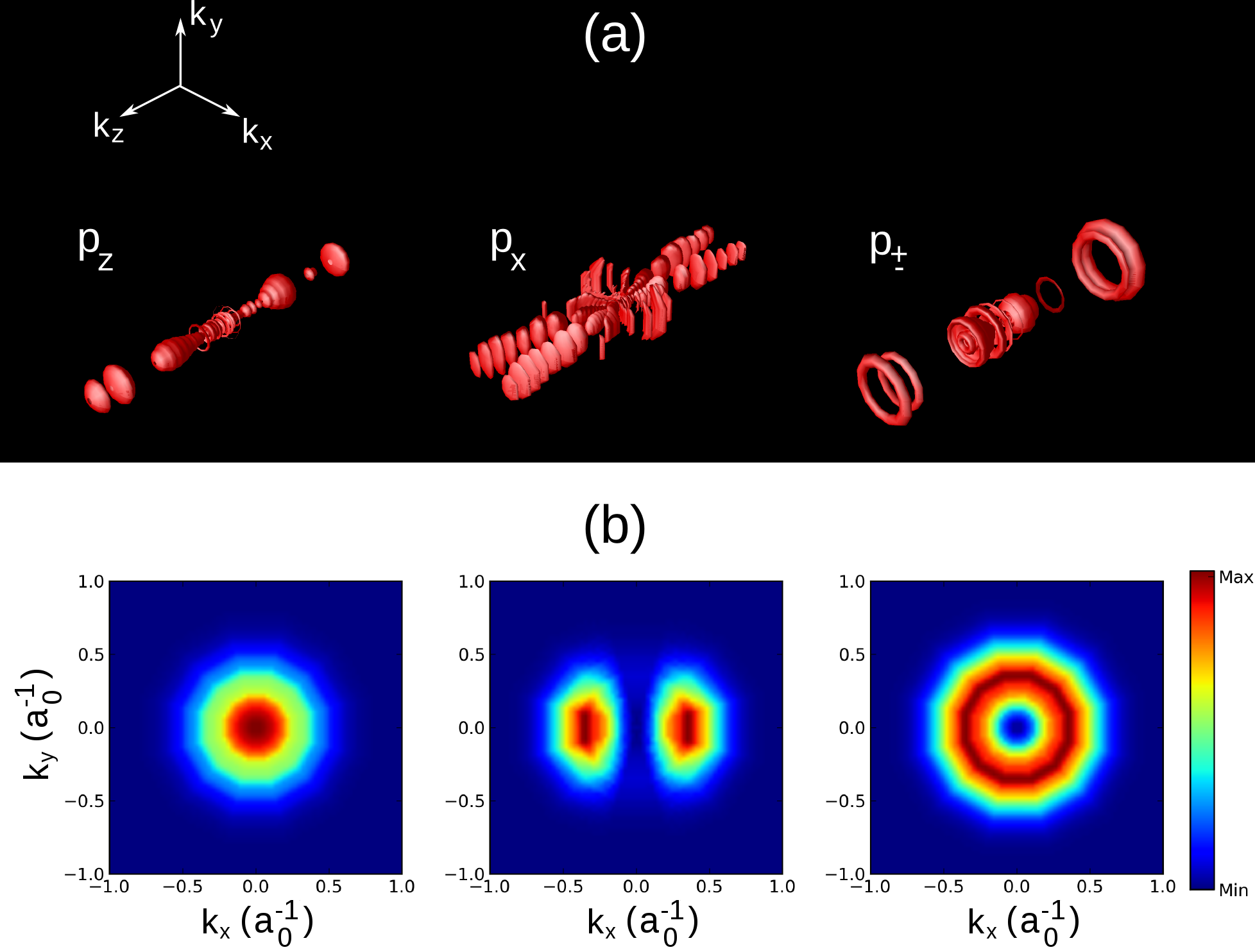}
\caption{(a) Isosurface plots of strong-field photoelectron spectra for the He$^+$ atom with initial
2p$_z$ (left), 2p$_x$ (middle) and 2p$_{\pm{}1}$ (right) states. The same linearly $z$-polarized strong-field from the main text was used. 
The isosurface values are 0.75 a$_0^3$, 0.2 a$_0^3$ and 0.75 a$_0^3$, respectively.
(b) Corresponding two-dimensional cuts in the k$_x$k$_y$-plane at k$_z$=2.1 $a_0^{-1}$.}
\label{Fig_initial_state_symmetry_ps}
\end{figure}

\begin{figure}[h]
\centering
\includegraphics[height=0.9\textheight]{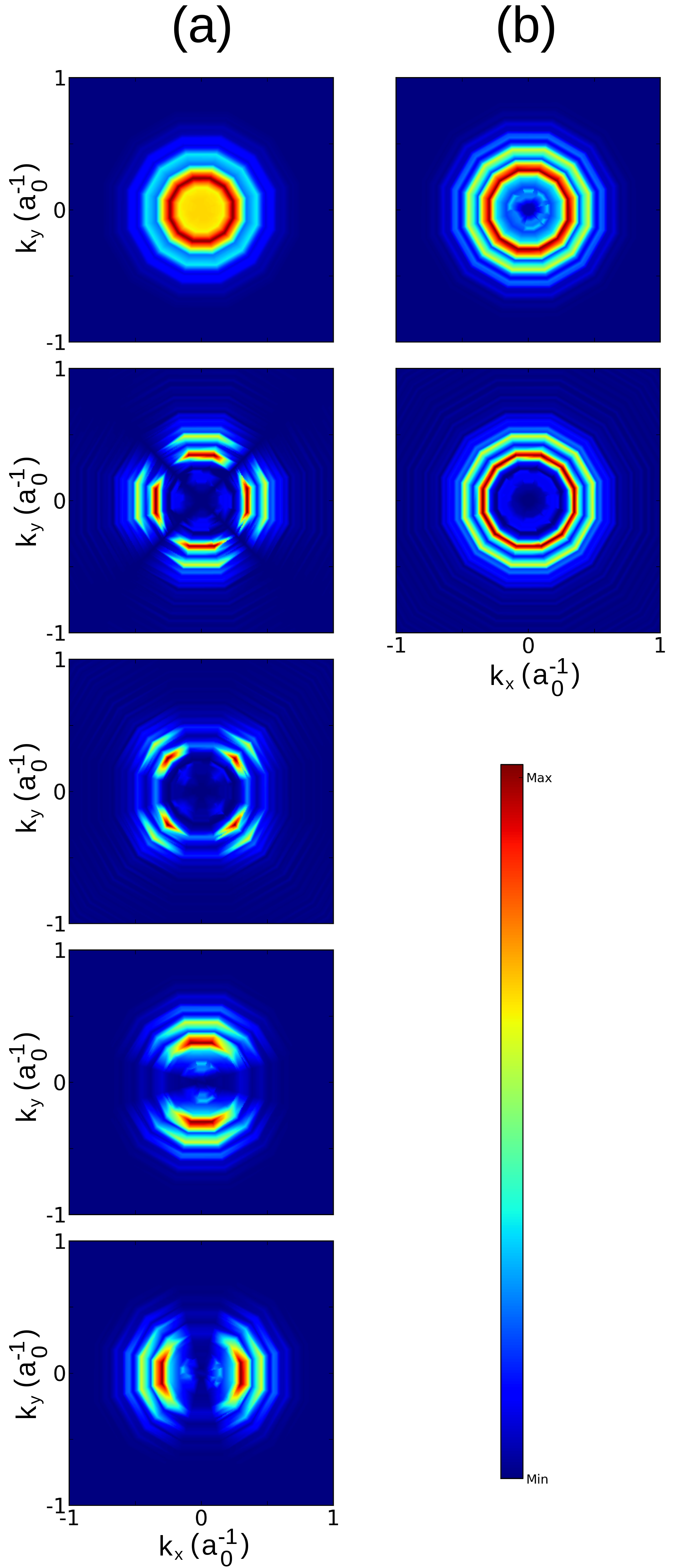}
\caption{Same as Figure \ref{Fig_initial_state_symmetry_ps}(b) for the He$^+$ atom initially in the (from top to bottom) 3d$_{z^2}$, 3d$_{x^2-y^2}$, 
3d$_{xy}$, 3d$_{yz}$, 3d$_{zx}$ state (a). Panel (b) shows the corresponding results for initial 3d$_{\pm{}1}$ (top) and 3d$_{\pm{}2}$ (bottom) states.}
\label{Fig_initial_state_symmetry_ds}
\end{figure}

Figure \ref{Fig_initial_state_symmetry_ps} adds ring-current carrying states to the discussed initial-state symmetries, see rightmost panels.
For comparison to the main text, the left panels of Fig. \ref{Fig_initial_state_symmetry_ps} show the corresponding strong-field photoelectron
spectra for the symmetric initial 2p$_z$ state and the middle panels for the anti-symmetric initial 2p$_x$ state already discussed in 
Figure 1 in the main text. Here we visualize the three-dimensional photoelectron spectrum by means of isosurfaces (Fig. \ref{Fig_initial_state_symmetry_ps}(a)),
and by two-dimensional cuts perpendicular to the laser-polarization direction ($z$-direction, Fig. \ref{Fig_initial_state_symmetry_ps}(b)). We nicely see how the initial-state
symmetries are mapped to the strong-field photoelectron spectrum. 
In particular, current-carrying initial states give rise to electronic vortex (doughnut) beams with intensity minimum in the center of the beam, which is mapped
to the photoelectron spectrum. Figure \ref{Fig_initial_state_symmetry_ds}(b) shows the corresponding two-dimensional cuts for initial 
ring-current carrying 3d$_{\pm{}1}$ (top) and 3d$_{\pm{}2}$ states (bottom). 
Light vortices are used in super-resolution light microscopy, see Ref. [26] in the main text. 
Here, we demonstrate atomic-scale electronic vortex beams, which may enhance the resolution in strong-field spectroscopies, giving rise to
attosecond, sub-\AA{}ngstr\"om super-resolution spectroscopy.

Moreover, the high-energy rescattering region of the strong-field photoelectron spectra carries background-free imprints of the initial-state symmetry 
in the planes transverse to the laser polarization direction, see Figs. \ref{Fig_initial_state_symmetry_ps} and \ref{Fig_initial_state_symmetry_ds}.
Part (a) of the latter Figure shows the corresponding planes for the five real d-orbitals (from top to bottom) 3d$_{z^2}$, 3d$_{x^2-y^2}$, 
3d$_{xy}$, 3d$_{yz}$, 3d$_{zx}$ as initial states.

\section{Transverse structures of rescattered photoelectrons in molecules}

\begin{figure}[h]
\centering
\includegraphics[width=0.9\textwidth]{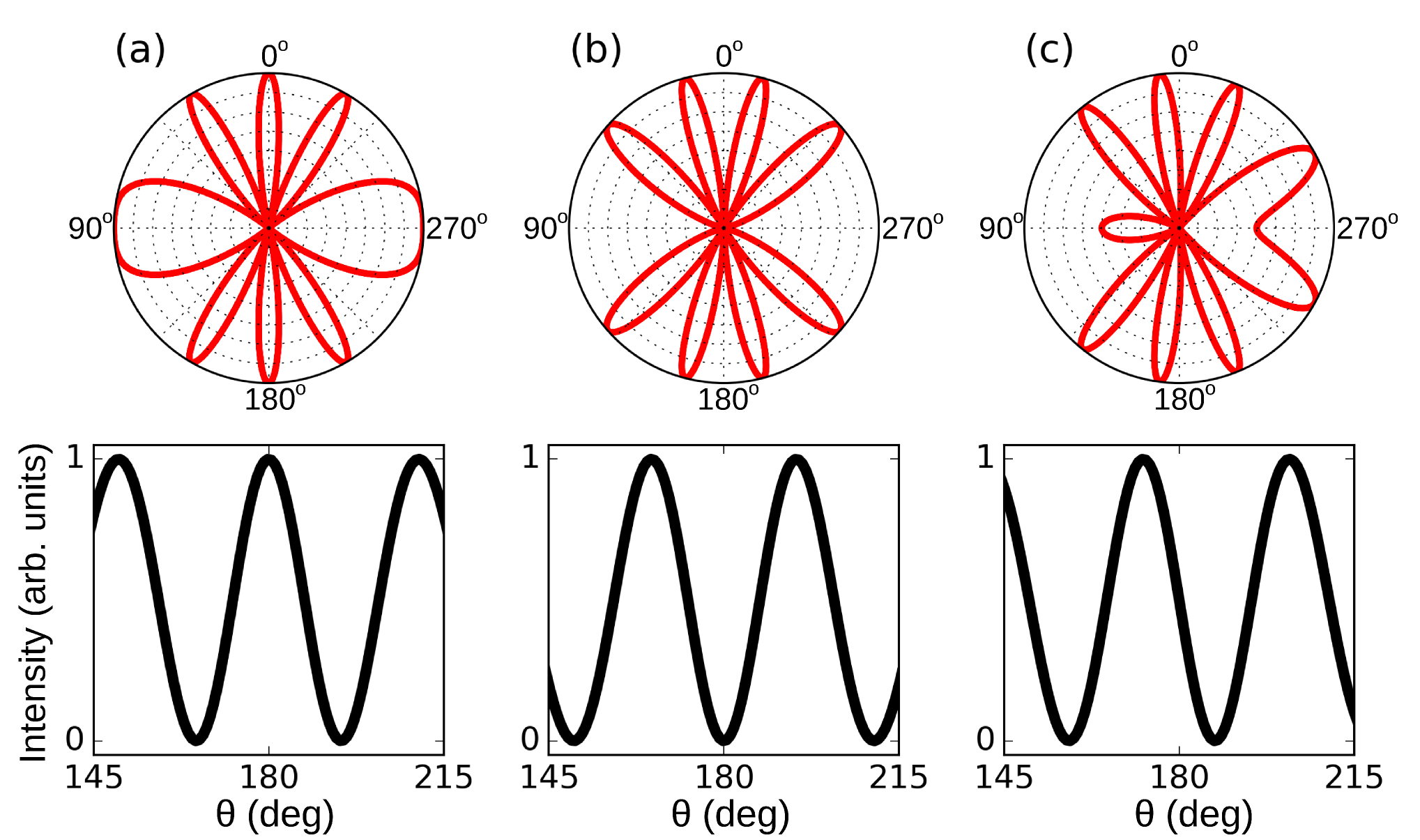}
\caption{Far-field double-slit diffraction pattern. The ratio of the slit separation
$d$ and the wavelength $\lambda$ of the incoming beam is $d/\lambda=2$.
The phase difference $\Phi$ between the two emitters is chosen to
be $\Phi=0$ (a), $\Phi=\pi$ (b) and $\Phi=\pi/2$ (c), see text
for details. The upper panels show the far-field diffraction patterns
for diffraction angels $0^{\circ}\leq\theta\leq360^{\circ}$, while
the lower panels show details in the range $145^{\circ}\leq\theta\leq215^{\circ}$. }
\label{Fig1_SI}
\end{figure}

Implications and opportunities of transverse structures of rescattered
photoelectrons in molecules may be illustrated by means of a double-slit \textit{Gedankenexperiment}. For simplicity, we consider the
two-dimensional case, where a monochromatic plane wave, with wavevector
$k=2\pi/\lambda$, illuminates two narrow slits separated by a
distance $d$ at normal incidence. 
Figure \ref{Fig1_SI}(a) shows
the corresponding far-field diffraction pattern, $I(\theta)\propto{\rm cos}(\pi\frac{d}{\lambda}{\rm sin}(\theta))^{2}$,
as function of the diffraction angle $\theta$ for $\frac{d}{\lambda}=2$.
The characteristic diffraction minima allow a straightforward determination
of the slit separation $d$. 

Figure \ref{Fig1_SI}(b) shows the analogous case for an incoming
wave with $\Phi=\pi$ phase jump between the slits, giving rise to
the far-field diffraction pattern $I(\theta)\propto{\rm cos}(\pi\frac{d}{\lambda}{\rm sin}(\theta)+\frac{1}{2}\Phi)^{2}$.
This case corresponds to strong-field ionization and rescattering
along a nodal plane as demonstrated in the main text. 
In particular, the returning electronic wavepacket can no longer
be described by an asymptotically-flat wavefront with well-defined return direction
with respect to the laser polarization direction, as is typically done in
state-of-the-art LIED experiments, \textcolor{black}{see e.g. Refs. [13,18-23] in the main text}.
Instead, the returning electronic wavepacket may be approximated by two trajectories with $\pi$
phase difference, see Fig. 1(a) (right panel) and discussion following eqn. (3) in the main text.
This scenario translates directly to molecules
due to the underlying symmetry. 

Finally, Figure \ref{Fig1_SI}(c)
shows the diffraction pattern for a phase difference of $\Phi=\pi/2$
between the two emitters. This case corresponds to imperfect alignment
between the laser polarization axis and the symmetry element of the
initial state, demonstrated in the main text, see Figures 2 and 3.
Here, the returning electronic wavepacket is composed of different return
directions which need to be added coherently, see eqn. (4) in the main text. Also,
these return directions depend sensitively on the specific quantum system as 
shown in the main text, c.f. Figs. 2 and 3. 
\begin{figure}[t]
\centering
\includegraphics[width=1\textwidth]{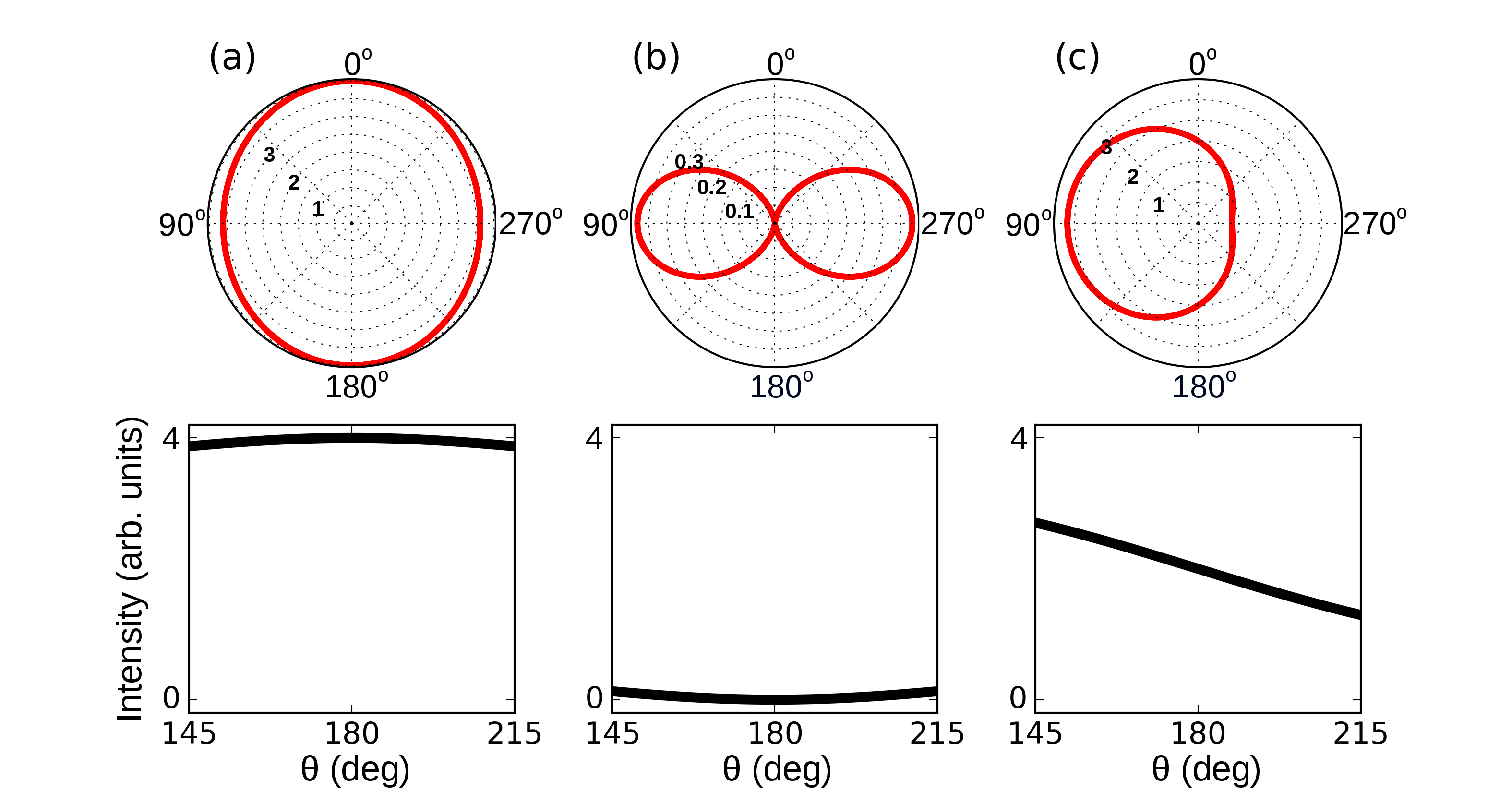}
\caption{Same as Figure \ref{Fig1_SI} for $d/\lambda=0.1$. }
\label{Fig2_SI}
\end{figure}

The demonstrated robust transverse phase structures of the
rescattered photoelectron are not only crucial in the interpretation of state-of-the-art
strong-field rescattering experiments but also present opportunities for strong-field
imaging.
Figure \ref{Fig2_SI} illustrates, in addition to attosecond vortex beams (Appendix A), 
another way of how these transverse
structures may enable super-resolution
strong-field imaging. Again we stress that, while the main text treated
atoms, such transverse structures of rescattered photoelectrons will
also be present in molecular systems due to the underlying symmetry,
see also Refs. {[}12,25{]}. Specifically, analogous to super-resolution
light microscopy (Ref. {[}26{]}), structured rescattered photoelectrons
may enable structure determination beyond the diffraction limit. Figure
\ref{Fig2_SI}(a) shows the standard far-field double-slit diffraction
pattern far beyond the diffraction limit for a ratio $d/\lambda=0.1$. Besides
the absence of diffraction minima, the overall diffraction pattern
is rather unstructered, rendering structural determination with realistic
signal-to-noise ratios practically impossible. The same is true for
$\Phi=\pi$ (Figure \ref{Fig2_SI}(b)), i.e. a $\pi$ phase shift
between the emitters. On the other hand, for $\Phi=\pi/2$ (Figure
\ref{Fig2_SI}(c)) the pattern is much more structured, which is most
apparent close to the backscattering region around $\theta=180^{\circ}$
crucial to strong field imaging, Refs. [5,6,10,13-23] in the main text.
Specifically, the slope of the diffraction pattern in the backscattering
region is related to the slit separation $d$ through $I^{\prime}(\theta=0)\propto-{\rm cos}(\frac{1}{2}\Phi){\rm sin}(\frac{1}{2}\Phi)\frac{d}{\lambda}$.
Hence, with known phase difference $\Phi$ between the emitters, the
determination of the slit separation $d$ becomes in principle possible,
even much beyond the diffraction limit. On the other hand, a known slit separation (known internuclear distance)
allows the determination of the phase-gradient of the returning electronic wavepacket.

Also, detailed analysis of the tilt
of the diffraction pattern as function of the misalignment angle between
the laser polarisation and initial state symmetry elements (see Figures
2(c) and 3(b) in the main text) may enable the determination of the
phase structure of the recolliding electronic wavepacket. The
corresponding shift of the holographic fringes may equally aid in the characterisation
of the structure of the continuum wavepacket, see also Ref. {[}12{]}. 

%
%

\section{Focal averaging}
\begin{figure}[t]
\centering
\includegraphics[height=0.75\textheight]{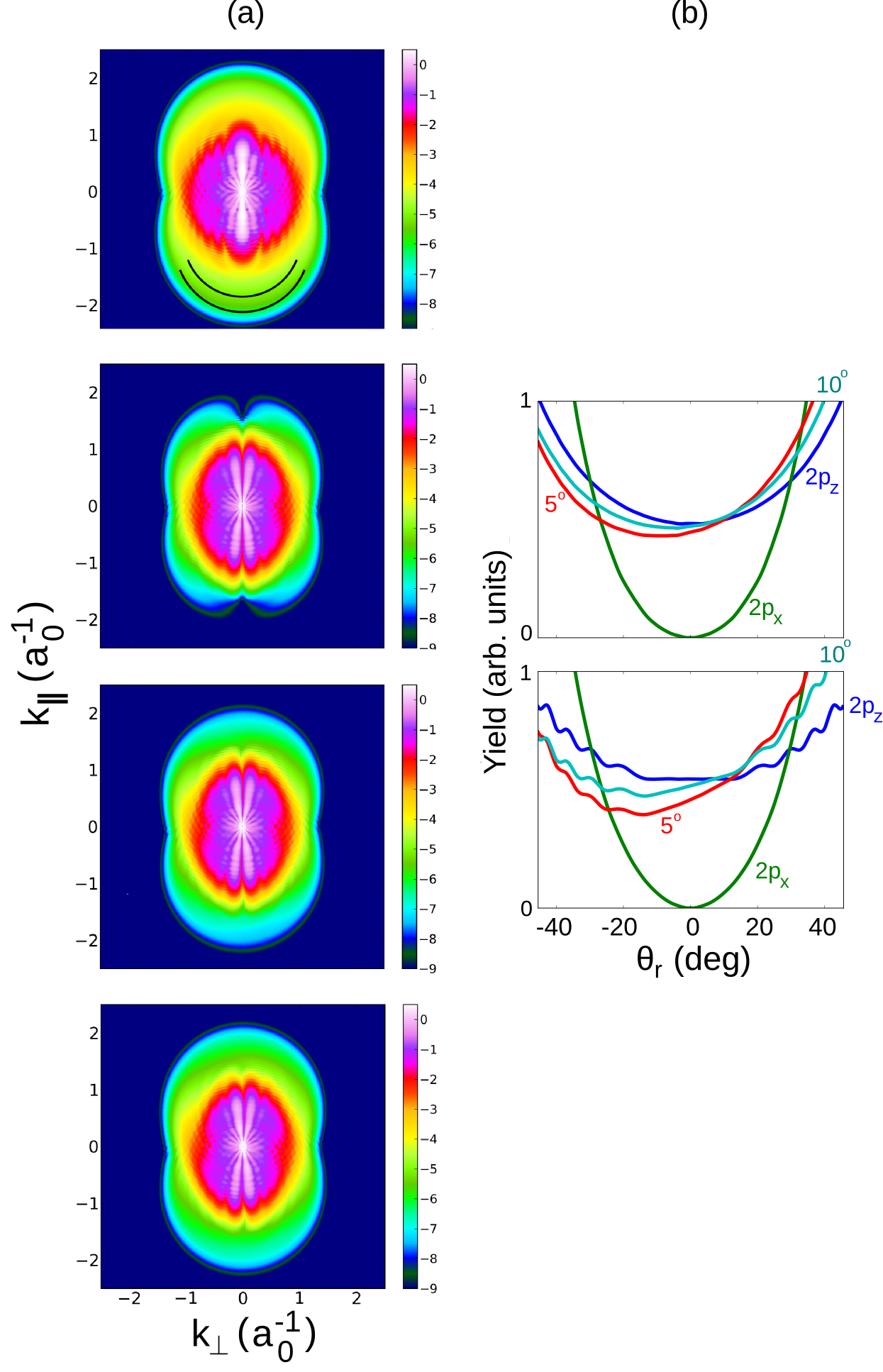}
\caption{(a) Focal-averaged strong-field photoelectron spectra for the He$^+$ atom at the 
peak intensity of $10^{14}$ W/cm$^2$ for (from top to bottom): symmetric initial $2p_z$, anti-symmetric initial $2p_x$
as well as ``misaligned'' He$^+$ using misalignment angles of $\alpha=5^{\circ}$ and $\alpha=10^{\circ}$, see main text for details.
(b) Corresponding angle-resolved photoelectron yields along the 10U$_{\rm{p}}$ (top) and the 8U$_{\rm{p}}$ (bottom) recollision circles sketched in part (a) for
initial $2p_z$ (blue), $2p_x$ (green) as well as $\alpha=5^{\circ}$ (red)
and $\alpha=10^{\circ}$ (cyan), see text for further details.}
\label{Fig3_SI}
\end{figure}
Figure \ref{Fig3_SI} explores the effect of focal averaging for the He$^+$ atom.
We assume a Gaussian beam for which the volume of the focal
spot is given by:
\begin{align*}
   & V  = V_0 ( y^3 + 6(y-\rm{atan}(y)) ) \qquad & \rm{with} \\
   & V_0 = (2\pi/9) w_0^2 * Z_r \qquad & \rm{and}\\
   & y  = \sqrt{I_{\rm{max}}/I_0 - 1}
\end{align*}
where $I_{\rm{max}}$ is the peak intensity, $w_0$ is the beam waist, and $Z_r$ is the Rayleigh range.
Because we are not interested in the absolute numbers, we set $V_0$ to 1 from now on.

Figure \ref{Fig3_SI}(a) shows focal averaged strong-field photoelectron spectra for a peak intensity of
$I_{\rm{max}}=10^{14}$ W/cm$^2$. The remaining laser parameters correspond to the values used in
the main text. 
We stop volume integration at the isosurface corresponding to $I_0=10^{13}$ W/cm$^2$, where strong-field ionization
becomes negligible.
Four different initial
states are used, namely (from top to bottom) symmetric initial $2p_z$, anti-symmetric initial $2p_x$
as well as ``misaligned'' He$^+$ using misalignment angles of $\alpha=5^{\circ}$ and $\alpha=10^{\circ}$.
We observe the same suppression of the photoelectron signal in the ``holographic'' region for zero perpendicular final momentum
for the anti-symmetric initial state, which is stable with respect to ``misalignment'', c.f. Figure 2 in 
main text.

Figure \ref{Fig3_SI}(b) explores the effect of focal averaging on the ``LIED'' region. 
The top panel shows the angle-resolved strong-field photoelectron yields for the four orientations from Fig. \ref{Fig3_SI}(a),
i.e. initial symmetric $2p_z$ (blue), anti-symmetric $2p_x$ (green) as well as results for misalignment angles of $\alpha=5^{\circ}$ (red)
and $\alpha=10^{\circ}$ (cyan). The recollision circle is the same as in Figure 1(b) in the main text (``10U$_{\rm{p}}$'' recollision circle, see also
outermost circle in Fig. \ref{Fig3_SI}(a)). The lower panel of Fig. \ref{Fig3_SI}(b) shows results along the ``inner'' recollision 
circle sketched in Fig. \ref{Fig3_SI}(a). Again, we observe that the effects discussed in the main text for a single peak intensity
of $10^{14}$ W/cm$^2$ are stable with respect to focal averaging.

%
%

\section{Peak laser intensity}
\begin{figure}[t] 
\centering
\includegraphics[height=0.9\textheight]{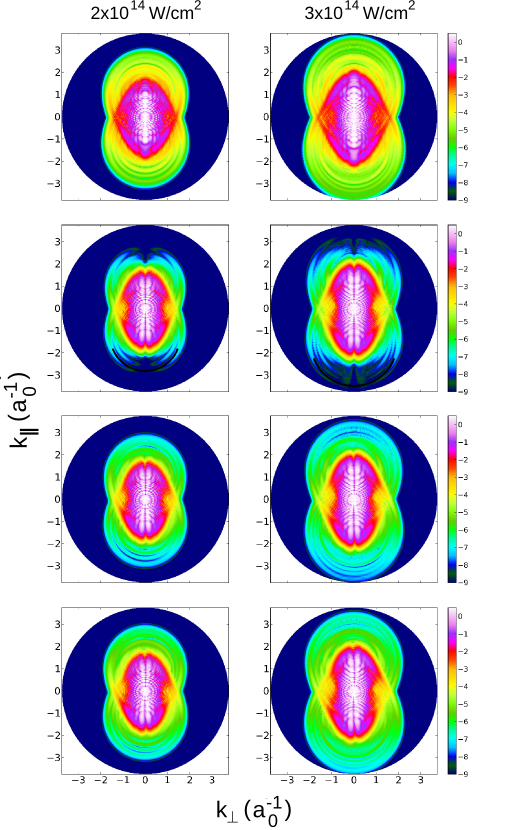}
\caption{Strong-field photoelectron spectra for the He$^+$ atom for peak laser intensities of $I_{\rm{max}}=2\times{}10^{14}$ W/cm$^2$ (left) and
$I_{\rm{max}}=3\times{}10^{14}$ W/cm$^2$ (right). The same four different initial states and remaining laser parameters as in Figure \ref{Fig3_SI} are used.}
\label{Fig4_SI}
\end{figure}

\begin{figure}[t]
\centering
\includegraphics[height=0.5\textheight]{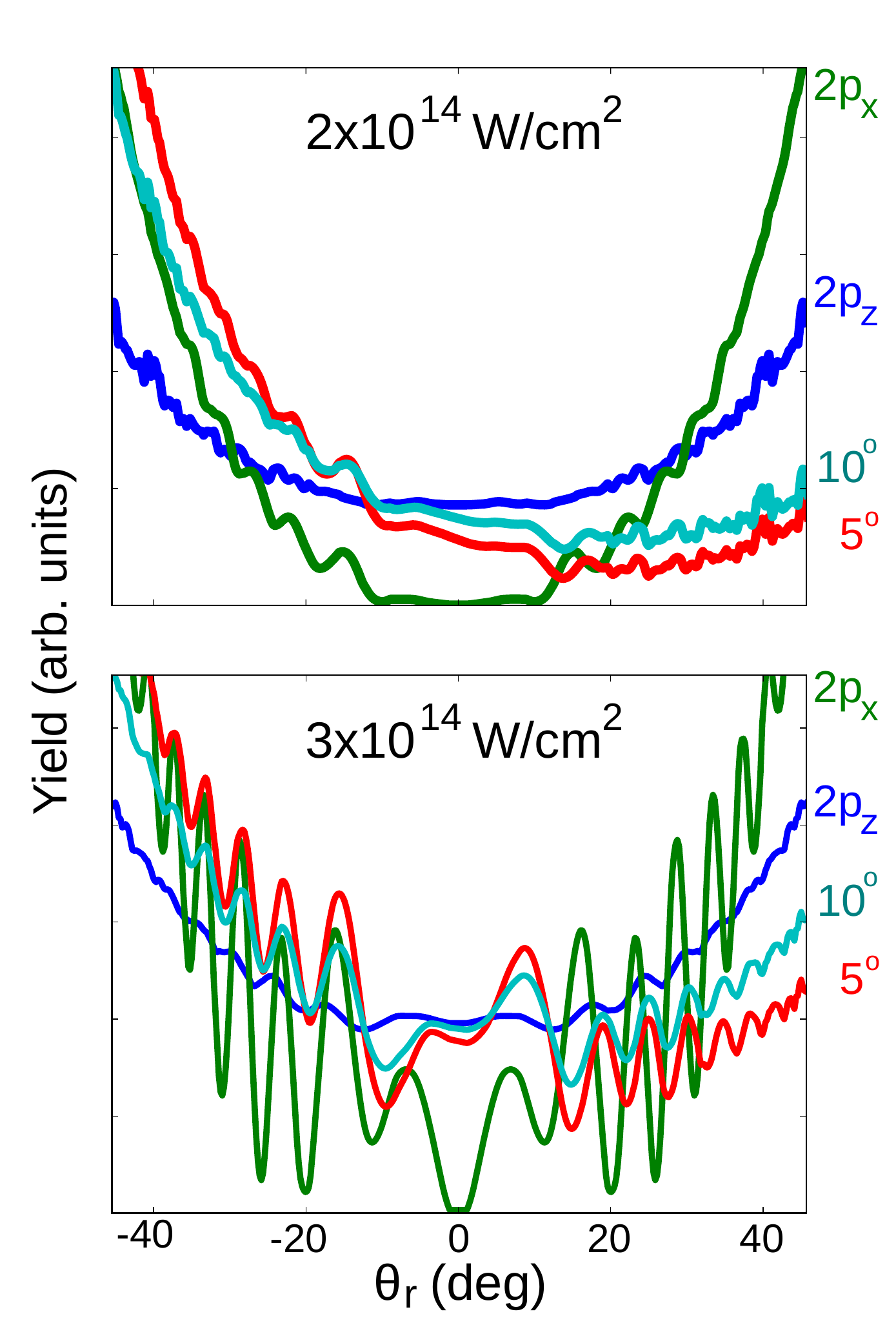}
\caption{Angle-resolved photoelectron yields for the He$^+$ atom along the outer recollision circles sketched in Fig. \ref{Fig4_SI} as black lines for the four different initial states 
and the two peak laser intensities from Figure \ref{Fig4_SI}. The color coding is the same as used in Figure \ref{Fig3_SI}(b).}
\label{Fig5_SI}
\end{figure}

Figures \ref{Fig4_SI} and \ref{Fig5_SI} explore the robustness of the observed
features as function of the peak laser intensity. Figure \ref{Fig4_SI} shows 
strong-field photoelectron spectra for a peak intensity of $I_{\rm{max}}=2\times{}10^{14}$ W/cm$^2$ (left panels)
and a peak intensity of $I_{\rm{max}}=3\times{}10^{14}$ W/cm$^2$ (right panels). Angular momentum channels 
up to $L\leq{}80$, $|M|\leq{}80$ are included for $I_{\rm{max}}=2\times{}10^{14}$ W/cm$^2$ and channels up to
$L\leq{}120$, $|M|\leq{}120$ are included for $I_{\rm{max}}=3\times{}10^{14}$ W/cm$^2$. The remaining simulation parameters
are the same as used in the main text.
The same initial states from
Figure \ref{Fig3_SI} are used. Again, the strong-field photoelectron spectra 
are suppressed in the ``holographic'' region close to zero perpendicular final momentum ($k_{\perp}=0$)
for the anti-symmetric initial state, which is stable with respect to ``misalignment''.

Finally, Figure \ref{Fig5_SI} shows the angle-resolved photoelectron yields along the outer recollision
circles shown as black lines in the photoelectron spectra for the anti-symmetric initial $2p_x$ states in Figure \ref{Fig4_SI}.
The same color coding from Figure \ref{Fig3_SI}(b) is used. We observe again that the transverse structure of the returning
electronic wavepacket strongly affects strong-field rescattering. The effect is robust to field misalignments with respect 
to the symmetry plane. 

\bibliography{recollision}

\end{document}